\documentclass[%
 reprint,
amsmath,amssymb,
apl,
prb
]{revtex4-2}

\usepackage{graphicx}
\usepackage{dcolumn}
\usepackage{bm}
\usepackage{courier}
\usepackage{amsmath,amssymb}
\usepackage{mathrsfs}
\usepackage{bm}
\usepackage{supertabular}
\usepackage{float}
\usepackage{soul}
\usepackage{color}
\usepackage{natbib}
\usepackage{placeins}
\usepackage{array}
\usepackage{multirow}
\usepackage{afterpage}
\usepackage{lipsum}
\usepackage[utf8]{inputenc}
\usepackage{csquotes}
\usepackage{braket}

\begin{document}

\preprint{APS/123-QED}

\title{Role of Dimensionality on Excitonic Properties of BiSeI using Many-body Perturbative Approaches}

\author{Sanchi Monga}
 \email{sanchi@physics.iitd.ac.in[SM]}
\author{Saswata Bhattacharya}
 \email{saswata@physics.iitd.ac.in [SB]}
\affiliation{Department of Physics, Indian Institute of Technology Delhi, New Delhi 110016, India}
\begin{abstract}

The mechanical exfoliation of two-dimensional materials has sparked significant interest in the study of low-dimensional structures. In this work, we investigate the bulk and low-dimensional derivatives of BiSeI, a quasi-one-dimensional anisotropic crystal known for its remarkable stability and novel electronic properties. Using the density functional theory and many-body perturbation theory, we examine the influence of dimensionality on their electronic, optical, and excitonic properties. Quasi-particle $\mathit{G_0W_0}$ calculations reveal a significant increase in the band gap with a decrease in dimensionality, driven by quantum confinement effects and reduced dielectric screening. By solving the Bethe-Salpeter equation, we identify a transition from weakly bound Wannier-Mott excitons in bulk BiSeI to strongly bound excitons its low-dimensional forms. These structures feature band gaps spanning the infrared to the visible spectrum and exhibit large exciton binding energies, making them promising for next-generation optoelectronics and excitonic applications. Our findings provide a theoretical foundation for future experimental studies on BiSeI and its low-dimensional counterparts.
\end{abstract}

\maketitle

\section*{Introduction}
\begin{figure*}[t]
    \centering
    \includegraphics[width=0.9\textwidth]{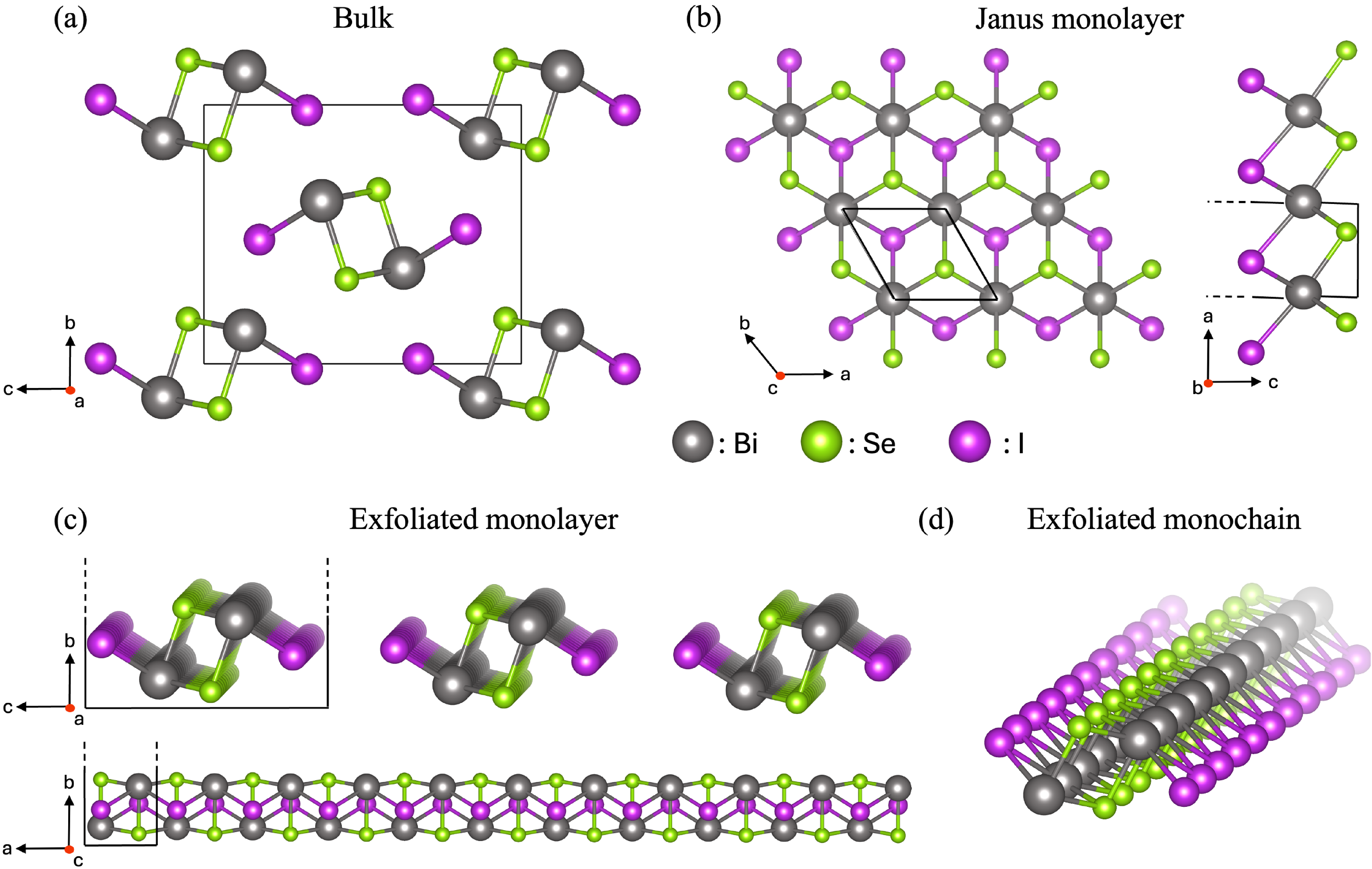}
    \caption{Crystal structure of (a) Bulk, (b) Janus monolayer, (c) exfoliated monolayer, and (d) exfoliated monochain BiSeI. Black box denotes unit cell of the respective structures.}
    \label{structure}
\end{figure*}
Hybrid organic-inorganic lead halide perovskites (LHPs), represented by the general formula ABX$_3$ (A = CH$_3$NH$_3$$^+$, CH(NH$_2$)$_2$$^+$, or mixed A-cations; B = Pb$^{2+}$; X = I$^{-}$, Br$^{-}$, Cl$^{-}$), have emerged as leading contenders in optoelectronic and solar light harvesting applications due to their exceptional electronic and optical properties \cite{Green2014,Yin2015,kim2020high,yu2020miscellaneous}. Over the last decade, LHP-based solar cells have demonstrated remarkable progress, with power conversion efficiencies increasing from 3.9 $\%$ in 2009 to more than 25 $\%$ \cite{nrel}. This outstanding performance is primarily attributed to the Pb ns$^2$ lone pair, which enables the formation of highly dispersive valence and conduction bands, along with enhanced dielectric screening from the cross-band hybridization resulting in high charge carrier mobility and defect tolerance. Despite their remarkable optoelectronic properties, significant challenges remain, including lead toxicity and poor environmental stability, which limit their commercial viability ~\cite{schileo2021lead, ju2018toward}. These concerns have driven efforts to explore alternative materials with similar lone-pair characteristics, particularly those incorporating heavy post-transition metals such as bismuth (Bi) or antimony (Sb).

Chalcohalide materials, represented as MChX (M: metal cation, Ch: divalent chalcogen anion - S/Se, and X: monovalent halogen - Cl/Br/I), are an emerging class of materials due to their widespread availability, non-toxicity, and reduced environmental impact \cite{ghorpade2022emerging}. A significant advantage of chalcohalides over LHPs lies in their superior structural stability, which arises from the covalent character of the M-Ch bonds, in contrast to predominantly ionic bonds found in LHPs. Among these, bismuth selenoiodide (BiSeI), a heavy pnictogen chalcohalide, exhibits remarkable electronic and optical properties akin to LHPs including a suitable band gap of 1.29 eV, small carrier effective masses, high dielectric constants, and strong optical absorption \cite{shi2016bismuth,shin1994optical, xiao2019centimeter, bai2022growth,shi2016bismuth}.

BiSeI is a quasi-one-dimensional anisotropic material, characterized by [BiSeI]$_\infty$ chains held together through weak van der Waals (vdW) interactions, resulting in a distinctive chain-like morphology (see Fig.~\ref{structure}a). This unique crystal structure makes BiSeI a promising candidate for generating low-dimensional systems via exfoliation or mechanical cleavage \cite{nicolosi2013liquid,coleman2011two,novoselov2005two,xu20232d,hu2023mixed,peng20181d}. Low-dimensional systems exhibit properties distinct from their bulk counterparts, primarily due to strong geometric confinement and reduced dielectric screening. These effects give rise to the formation of tightly bound electron-hole pairs (excitons) that remain stable even at room temperature \cite{komsa2012effects,raja2017coulomb,xiao2017excitons}. The optical properties of low-dimensional materials are predominantly governed by these strongly bound excitons, which can be leveraged in a range of optoelectronic applications, including photovoltaic solar cells, light-emitting diodes, and photodetectors \cite{wang2012electronics,tan20202d,mueller2018exciton}. Furthermore, dimensional reduction offers novel opportunities, such as blue shifting of the band gap due to quantum confinement effects enabling tunable light emission without altering the material's composition, which could otherwise introduce deep trap states. Despite these advantages, low-dimensional chalcohalides, including BiSeI, remain underexplored compared to their bulk counterparts.

In this study, we delve into the low-dimensional structures, viz. two-dimensional (2D) monolayer and one-dimensional (1D) monochain exfoliated from bulk BiSeI along with the theoretically predicted BiSeI Janus monolayer \cite{riis2019classifying,li2023polarization,varjovi2021first}. We start with a comparative analysis of the electronic properties of these materials obtained using the state-of-the-art density functional theory (DFT) and \textit{GW} approximation within the many-body perturbation theory (MBPT) \cite{jiang2012electronic,fuchs2008efficient}. Since \textit{GW} level of theory does not incorporate excitonic features, we solve the Bethe-Salpeter equation (BSE) \cite{luo2018efficient} to obtain the optical absorption spectra and excitonic properties of these materials. This limited research highlights the untapped potential of BiSeI and underscores the need for further investigation to fully exploit its unique properties.
\begin{table*}
    \centering
    \begin{tabular}{lccccccccc}
    \hline \hline
         && Bulk && Janus monolayer && Exfoliated monolayer && Exfoliated Monochain \\
         \hline
        a (\text{\r{A}}) && 4.18 && 4.19 && 4.18 && 4.18\\
        b (\text{\r{A}}) && 8.63 && 4.19 && - && - \\
        c (\text{\r{A}}) && 10.64 && - && 10.64 && - \\
        \textit{GW} bands && 800 && 200 && 400 && 400 \\
        \textit{GW} dielectric cutoff && 8 Ry && 8 Ry && 8 Ry && 8 Ry \\
        Static $\epsilon (\omega=0)$ bands && 800 && 200 && 400 && 400 \\
        BSE bands && 188-198 && 41-56 && 91-102 && 91-102 \\
        $\mathit{k/q}$-grid && 8$\times$3$\times$4 && 24$\times$24$\times$1 && 15$\times$1$\times$6 && 40$\times$1$\times$1 \\
        \hline \hline
    \end{tabular}
    \caption{Computational settings for evaluating the electronic and optical characteristics of bulk, Janus monolayer, exfoliated monolayer and monochain BiSeI, incorporating spin-orbit coupling.}
    \label{params}
\end{table*}

\section*{Computational Details}

The first-principles calculations within the DFT framework are performed using the Vienna \textit{ab initio} Simulation Package (VASP) \cite{kresse1993ab,kresse1994ab,kresse1996efficiency,kresse1996efficient} and the Quantum Espresso (QE) suite \cite{giannozzi2009quantum}.
To incorporate interactions between valence and core electrons, fully relativistic projector augmented wave (PAW) \cite{kresse1999ultrasoft} and norm-conserving pseudopotentials \cite{van2018pseudodojo} are used for VASP and QE, respectively. The semicore 4\textit{d} and 5\textit{d} states for iodine and bismuth are explicitly included in the valence electrons due to their importance in the electronic structure, as discussed in the literature \cite{scherpelz2016implementation,giustino2014materials}. Geometry optimization is performed with the VASP code using the PBEsol \cite{perdew2008restoring} functional, a revised version of PBE \cite{perdew1996generalized} for solids. It accurately predicts lattice constants for systems with vdW interactions without explicitly accounting for them \cite{ganose2016relativistic,bjorkman2012van}. Bulk BiSeI is lattice optimized until the residual forces on the atoms are less than 10$^{-3}$ eV/\text{\AA}, with a plane wave kinetic energy cut-off of 500 eV. The relaxed lattice parameters (see Table ~\ref{structure}) closely align with the experimental values, with an error under 1 $\%$ \cite{braun2000bismuth}. This optimized bulk structure is used to exfoliate low-dimensional structures, followed by relaxation of their atomic positions. A 20 \text{\AA} vacuum is applied in low-dimensional structures to prevent fictitious interactions between periodic images. In order to determine the dynamical stability of these materials, we compute phonon band structures using the Phonopy package \cite{togo2008first,togo2015first}.

Since DFT underestimates the electronic band gap \cite{reining2018gw,onida2002electronic}, we perform single-shot \textit{GW} ($\mathit{G_0W_0}$) \cite{stan2009levels} to calculate the electronic band structures using the plasmon-pole approximation for the dynamical dependence of the dielectric function \cite{rojas1995space}. $\mathit{G_0W_0}$ does not include electron-hole interactions, neglecting excitonic effects. In order to incorporate the exciton formation and determine the optical spectra, we solve the BSE
within the Tamm-Dancoff approximation \cite{strinati1988application}. $\mathit{G_0W_0}$ and BSE, within the framework of MBPT, are performed using the YAMBO package \cite{marini2009yambo,sangalli2019many,marsili2021spinorial}. QE is used for the starting point ground state electronic structure calculations with a plane-wave kinetic energy cut-off of 40 Ry. The truncated Coulomb potential with slab and cylindrical geometry is used for the 2D and 1D structures to avoid interaction between repeated images along the non-periodic directions \cite{ismail2006truncation}. In order to speed up the convergence of electronic Green's function with respect to the number of unoccupied bands and $\mathit{k}$-grid, we use the Bruneval-Gonze terminator \cite{bruneval2008accurate} and the random integration method (RIM) \cite{guandalini2023efficient}. The number of bands, dielectric cut-off, and \textit{k}-grid used to calculate the dielectric matrix in \textit{GW}/BSE calculations are listed in Table ~\ref{params}.

\section*{Results and Discussion}

\subsection*{Crystal Structure and stability}
Bulk BiSeI crystallizes in an orthorhombic structure with the \textit{Pnma} space group, as shown in Fig.~\ref{structure}a. The crystal structure consists of [BiSeI]$\infty$ parallel ribbons aligned along the \textit{a}-axis, held together by weak vdW interactions. Bi and Se atoms are connected by covalent bonds, while I atoms form ionic bonds with the covalent Bi-Se bridge. The weak vdW interactions between the ribbons make it a strongly quasi-1D structure \cite{braun2000bismuth}. To evaluate the feasibility of exfoliating the 2D monolayer and the 1D monochain, we compute their exfoliation energies using the expression:
\[E_{exf} = E_{nD}/N_{nD} - E_{bulk}/N_{bulk}\]
where $E_{nD/bulk}$ and $N_{nD/bulk}$ represent the total energies and number of atoms in n-dimensional or bulk structures, respectively \cite{zhang20231d}. The estimated $E_{exf}$ values for the 2D monolayer and the 1D monochain are 200 meV and 212 meV. These values are lower than those of some experimentally exfoliated structures, indicating their experimental viability \cite{andharia2018exfoliation,choudhary2017high,zhang20231d}. Fig.~\ref{structure}c,d illustrates the exfoliated 2D monolayer and 1D monochain structures, exhibiting \textit{P}2$_1$\textit{m} space group symmetry with vacuum along the \textit{b} and \textit{b} / \textit{c} axis, respectively. We also investigate the theoretically predicted Janus monolayer (Fig.~\ref{structure}b), which exhibits \textit{P}3\textit{m}1 space group symmetry. The stability of these low-dimensional structures is a critical aspect to address. To evaluate their dynamical stability, we compute the phonon dispersion of these materials as depicted in Fig. S1 of the Supplemental Material (SM). The absence of negative frequencies in the phonon band structure confirms their dynamical stability.

\subsection*{Electronic structure}
\begin{figure}
    \centering
    \includegraphics[width=0.5\textwidth]{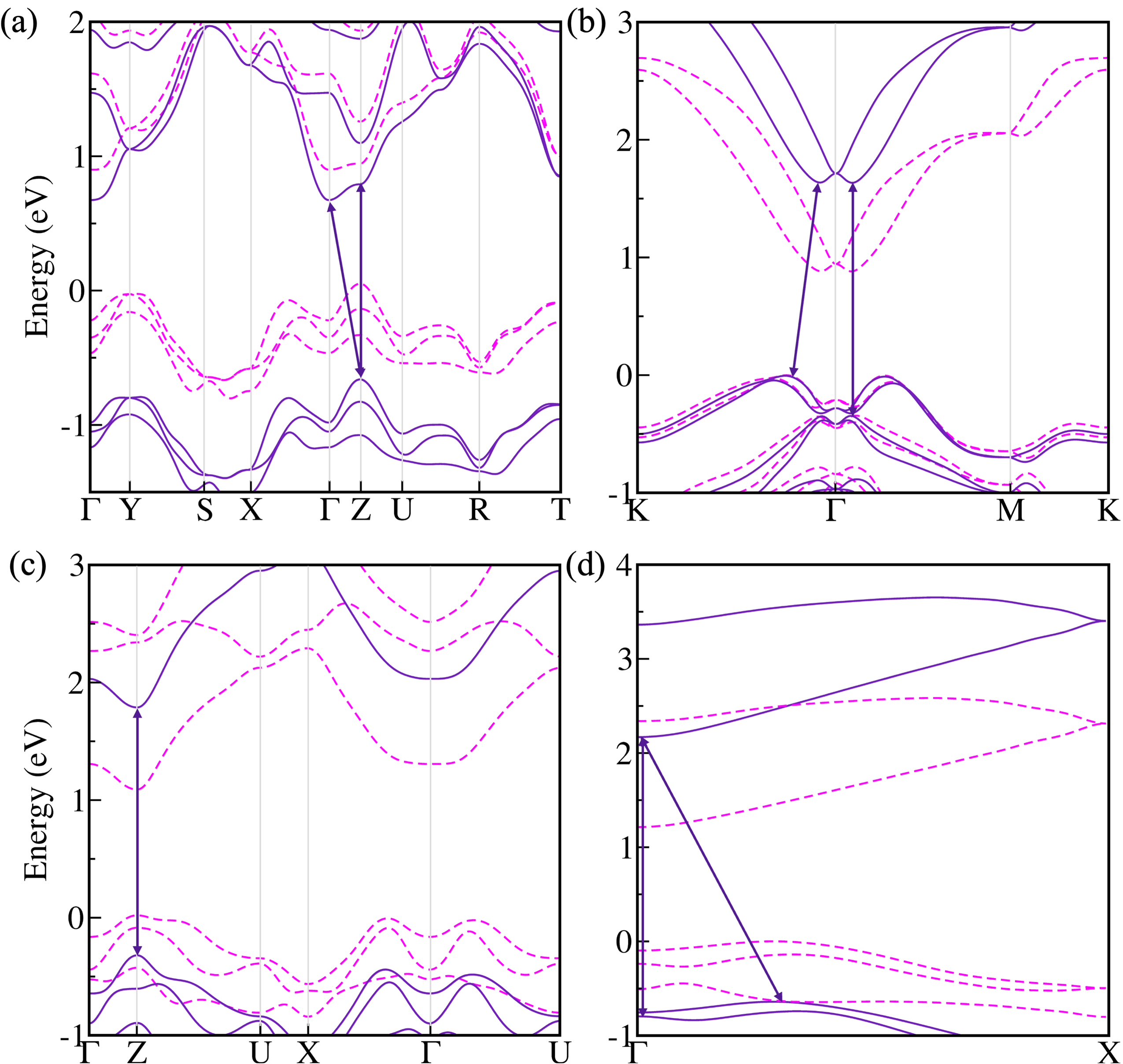}
    \caption{Band structure of (a) bulk, (b) Janus monolayer, (c) exfoliated monolayer and (d) exfoliated monochain of BiSeI computed using PBE (dashed) and $\mathit{G_0W_0@}$PBE (solid) methods, including spin-orbit coupling.}
    \label{band_structure}
\end{figure}
\begin{figure*}
    \centering
    \includegraphics[width=1\linewidth]{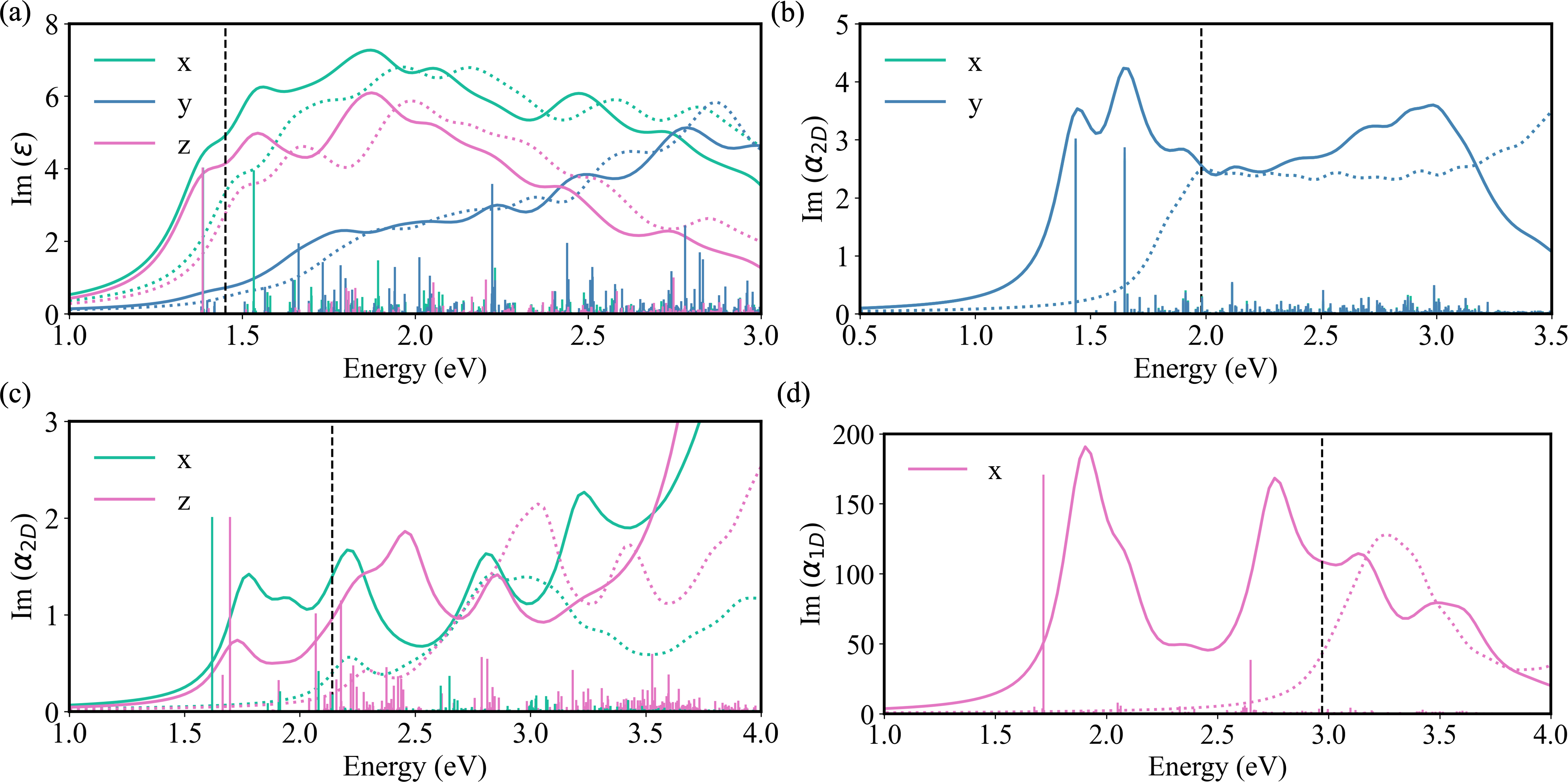}
    \caption{Absorption spectra for (a) Bulk, (b) Janus monolayer, (c) exfoliated monolayer, and (d) exfoliated monochain BiSeI, estimated using the $\mathit{G_0W_0}$ (dotted curve) and BSE (solid curve) methods, along each polarization direction. The vertical dashed black line indicates the $\mathit{G_0W_0}$ direct band gap and colored vertical lines indicate the oscillator strength for respective polarization directions.}
    \label{spectra}
\end{figure*}
To gain detailed insight into the electronic properties of Bulk BiSeI and its low-dimensional counterparts, we analyze their electronic band structures, obtained using the PBE and $\mathit{G_0W_0@}$PBE methods (Fig.~\ref{band_structure}). Arrows in the band structure indicate their lowest indirect and direct band gaps. As shown in Fig.~\ref{band_structure}, all structures, except the exfoliated 2D monolayer, are indirect band gap semiconductors. The band gap values are provided in Table \ref{gap}. Notably, the difference between the direct and indirect $\mathit{G_0W_0}$ band gap values is relatively small for the bulk (0.13 eV) and the 1D monochain (0.16 eV), whereas the 2D Janus monolayer exhibits a more substantial difference of 0.34 eV. 
In bulk BiSeI (Fig.~\ref{band_structure}a), the valence band maximum (VBM) occurs between the $\Gamma$ and Z high-symmetry points, while the conduction band minimum (CBM) is located at the $\Gamma$ point. Valence bands exhibit higher dispersion compared to conduction bands both parallel and perpendicular to the atomic chains, suggesting significant interchain coupling and relatively higher hole mobility compared to electron mobility. The conduction bands are however highly non-parabolic and possess maximum dispersion along the chain direction (\textit{a}-axis). Dimensional reduction leads to a notable increase in the band gap, which can be attributed to quantum confinement effects. For the 2D Janus monolayer (see Fig.~\ref{band_structure}b), an indirect band gap is observed with both VBM and CBM lying between the $\Gamma$ and K points. Furthermore, the difference between the two lowest indirect band gaps and the two lowest direct band gaps is merely 10 meV. However, for the exfoliated 2D monolayer (Fig.~\ref{band_structure}c), there is a direct band gap with both VBM and CBM located at the Z point. Similar to bulk, valence bands are more dispersive than conduction bands. Lastly, the exfoliated 1D monochain shows an indirect band gap, with VBM located between the $\Gamma$ and X points and CBM at the $\Gamma$ point. 
\begin{table}
    \centering
    \begin{tabular}{l c c c c c}
        \hline \hline
        & \multicolumn{2}{c}{DFT} && \multicolumn{2}{c}{\textbf{$\mathit{G_0W_0}$}} \\
        \cline{2-3}
        \cline{5-6}
        & ID & D && ID & D\\
        \hline
         Bulk & 0.85 & 0.89 && 1.33 & 1.46 (1.29$_e$\cite{Sadurni2024}) \\
         Janus monolayer & 0.88 & 1.15 && 1.64 & 1.98 \\
         Exfoliated monolayer & - & 1.07 && - & 2.14 \\
         Exfoliated monochain & 1.22 & 1.46 && 2.81 & 2.97 \\
         \hline \hline
    \end{tabular}
    \caption{Lowest indirect (ID) and direct (D) band gaps (in eV) computed using DFT (PBE) and beyond ($\mathit{G_0W_0@}$PBE) approaches, including spin-orbit coupling. Subscript \enquote{e} refers to the experimental value.}
    \label{gap}
\end{table}
Comparing the theoretical and experimental band gap values for bulk BiSeI, we find that the PBE functional significantly underestimates the band gap, a well-known limitation of this method. The $\mathit{G_0W_0}$ band structures, shown as solid curves in Fig.~\ref{band_structure}, yield band gap values in closer alignment with the experimental measurements \cite{Sadurni2024}. Although quasi-particle (QP) corrections increase the band gap, they do not alter the relative positions of the valence and conduction band edges, thus preserving the indirect or direct nature of the band gap. Additionally, QP corrections are more pronounced in low-dimensional structures due to reduced dielectric screening, which enhances electron-electron interactions. Given the presence of the heavy elements in BiSeI, spin-orbit coupling (SOC) plays a significant role in the electronic band structures, often shifting band edges by more than 0.5 eV (Table S1 of SM). To ensure accuracy, SOC effects have been included in all the calculations.

Fig. S2 shows the atomic orbital- and symmetry-projected partial density of states, calculated using the PBE functional. Across all dimensions, VBM is primarily determined by the hybridization of I-\textit{p} and Se-\textit{p} orbitals, with smaller contributions from Bi-\textit{s} and Bi-\textit{p} orbitals located deeper within the valence band. The interaction between Bi-\textit{s} and anionic-\textit{p} orbitals is expected to increase the valence band dispersion, akin to other halides with n\textit{s}$^2$ lone-pairs \cite{du2010enhanced,du2014efficient}, thereby reducing hole effective masses and enhancing hole mobility. On the other hand, CBM is predominantly characterized by Bi-\textit{p} orbitals with notable contributions from I-\textit{p} and Se-\textit{p} orbitals. This hybridization between cationic-\textit{p} and anionic-\textit{p} orbitals in the valence and conduction bands leads to substantial cross-band hybridization, imparting a mixed ionic-covalent character to the material. The mixed character, in turn, is known to impart strong lattice polarization and large dielectric constants \cite{du2010enhanced,du2014efficient}. 

\subsection*{Optical properties and excitonic effects}
The optical absorption spectra computed by solving the BSE on top of QP band structures are illustrated in Fig.~\ref{spectra}. The solid and dotted curves represent spectra with (BSE@\textit{G$_0$W$_0$}) and without (\textit{G$_0$W$_0$}) electron-hole interactions, respectively. \textit{G$_0$W$_0$} direct band gap is marked by the vertical dashed black line, while colored solid lines denote excitonic transitions. The first bright exciton with finite oscillator strength (here, greater than 5$\%$ of the maximum value) signifies the optical band gap of the material. The exciton binding energy is determined as the difference between the QP direct band gap and the exciton energy. For bulk materials, optical absorption is associated with the imaginary part of the macroscopic dielectric function, Im($\varepsilon$). However, in low-dimensional systems, the notion of macroscopic dielectric function becomes less meaningful. Instead, polarizability per unit area (for 2D systems) or per unit length (for 1D systems) is employed to represent optical absorption \cite{rohlfing2000electron,cervantes2024excitons,mella2024prediction}. Accordingly, we plot Im($\alpha_{2D/1D})$ against photon energy for monolayer and monochain configurations. 
\begin{figure}[h!]
    \centering
    \includegraphics[width=0.75\columnwidth]{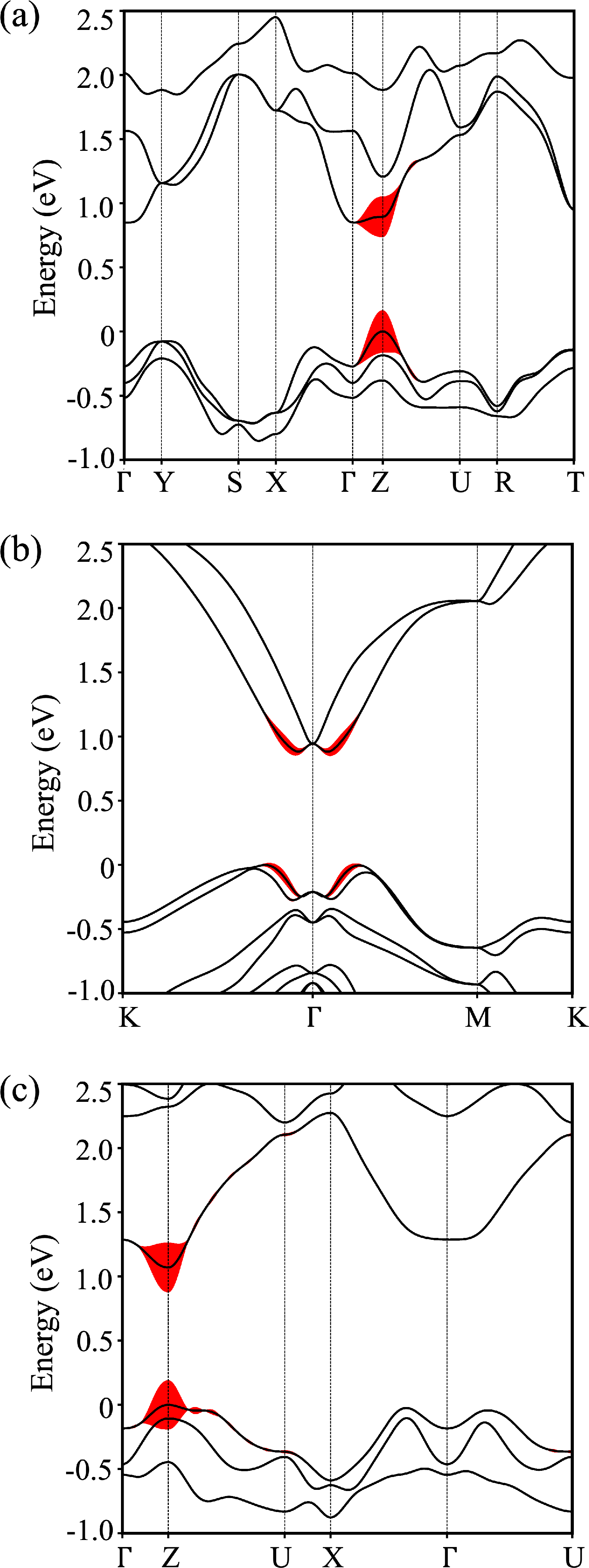}
    \caption{Exciton formation in (a) bulk, (b) Janus monolayer, and (c) exfoliated monolayer BiSeI. Exciton weights (depicted in red shade) are superimposed on the PBE+SOC one-electron band to visually inspect what kind of electron and hole states dominates the contribution to the first bright exciton.}
    \label{weights}
\end{figure}

Including electron-hole interactions induces a significant red shift in the optical spectra across all configurations. The presence of finite oscillator strength below the $\mathit{G_0W_0}$ direct band gap confirms the existence of bound excitons. In the bulk material, the first bright exciton peak appears at 1.39 eV (E $\parallel$ \textit{x} and \textit{z}) and 1.40 eV (E $\parallel$ \textit{y}), corresponding to exciton binding energies (E$_b$) of 70 meV, 70 meV, and 60 meV, respectively (see Fig.~\ref{spectra}a). Several dark excitons exist below the first bright exciton, with a bright-dark splitting ($\Delta$) of up to 10 meV. The optical absorption and oscillator strength of first bright excitons are significantly stronger along the atomic chain direction and the \textit{z}-axis compared to the \textit{y}-axis. This anisotropy in the absorption spectrum originates from the intrinsic geometric anisotropy of the crystal. Notably, the theoretically estimated optical band gap at 0 K is slightly larger than the experimental band gap (1.29 eV at 290 K), likely due to the exclusion of temperature-dependent effects in the theoretical calculations. In the isotropic 2D Janus monolayer, the first bright excitonic peak occurs at 1.64 eV, with a large binding energy of 0.34 eV, as depicted in Fig.~\ref{spectra}b. These excitons are doubly degenerate, both optically active, and no dark excitonic states are observed below the optical band gap. For the exfoliated monolayer (Fig.~\ref{spectra}c), the first bright excitonic peak appears at 1.66 eV (\textit{x}-polarization) and 1.62 eV (\textit{z}-polarization), with corresponding binding energies of 0.48 and 0.52 eV, respectively. 
\begin{figure*}
    \centering
    \includegraphics[width=1\linewidth]{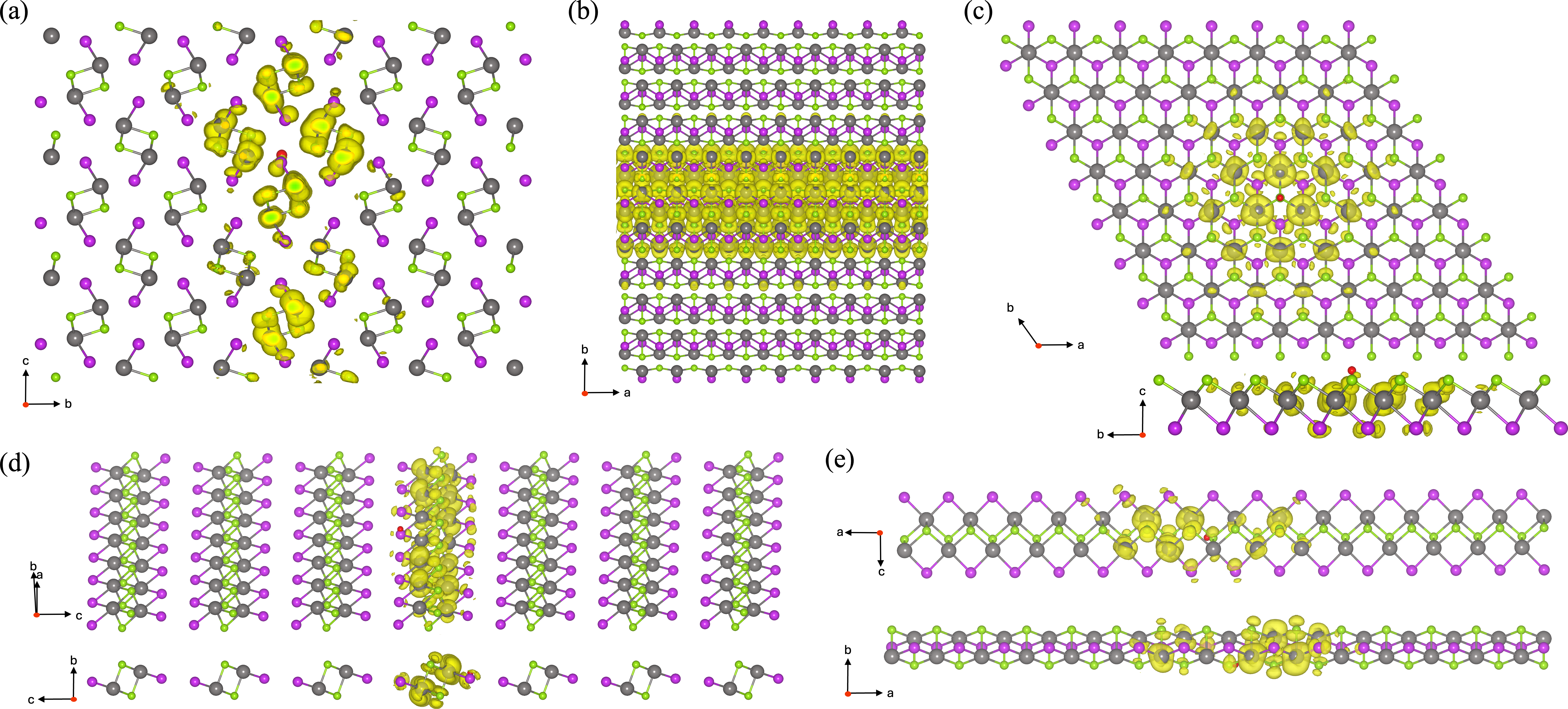}
    \caption{Top and side views of the wave functions of the first bright exciton for the bulk structure (a,b), Janus monolayer (c), exfoliated monolayer (d), and exfoliated monochain (e). The red dot represents the position of the hole, fixed at the atom contributing most significantly to the valence band edge. The isosurface value for visualizing the exciton wave function is set to 0.2 e/\AA$^3$. Grey, green and pink circles represent Bi, Se, and I atoms, respectively.}
    \label{wf}
\end{figure*}
The increased E$_b$ in the monolayer is attributed to reduced dielectric screening and enhanced quantum confinement effects characteristic of low-dimensional materials. Unlike the Janus monolayer, multiple dark excitons exist below the optical band gap, within $\Delta=$15 meV. In the 1D monochain (Fig.~\ref{spectra}d), the first bright exciton appears at 1.72 eV, with a remarkably high binding energy of 1.25 eV. This exciton is doubly degenerate, comprising one dark and one bright state. Additionally, five singlet dark excitons are observed below the optical band gap, with a bright-dark energy splitting of $\Delta=$0.29 meV.

For a detailed analysis of the optical transitions associated with the first bright exciton, we superimpose the exciton weights onto the DFT band structure (see Fig~\ref{weights}). This approach enables the identification of dominant \textit{k}-points in reciprocal space that contribute to the formation of this exciton \cite{kolos2022large,qu2023giant,kaur2025crossover}. The red shades in Fig~\ref{weights} represent the vertical electronic transitions from the direct band gaps that give rise to the first bright exciton in BiSeI. In bulk BiSeI, the first bright exciton arises from transitions between the VBM and CBM, predominantly localized at the Z high-symmetry point in reciprocal space and along the Z - $\Gamma$ and Z - U directions, as depicted in Fig~\ref{weights}a. In case of the Janus monolayer (Fig.~\ref{weights}b), this transition primarily occurs along the $\Gamma$ - K and $\Gamma$ - M paths, involving the VBM and CBM. Similar to bulk BiSeI, in the exfoliated monolayer shown in Fig~\ref{weights}c, the dominant contribution remains at the Z point and along the Z - $\Gamma$ and Z - U directions. However, the monolayer exhibits additional contribution from VBM-1 band, along with the VBM-to-CBM transition. For the exfoliated monochain, the first bright exciton, located at 1.72 eV, consists of transitions from VBM-2 and VBM-3 to CBM and CBM+1 at the $\Gamma$ point. However, accurately interpolating exciton weights onto the band structure necessitates a dense $\textit{k}$-grid, which exceeds our current computational scope. Consequently, while the relevant results have been successfully generated and analyzed, we have omitted the corresponding plot.

To visualize the first bright exciton in real space, we plot its isosurface onto the lattice of BiSeI, as shown in Fig~\ref{wf}. This exciton wavefunction is computed by fixing the hole at a position 1 $\text{\r{A}}$ above the unit cell, centered on the atom that predominantly contributes to the valence band maximum. The spatial extent of the isosurface provides insight into the exciton character, distinguishing between Wannier-Mott and Frenkel excitons. The wave function of the first bright exciton in bulk BiSeI, illustrated in Fig.~\ref{wf}a,b, reveals the spatial distribution of the electron with the hole localized on an iodine atom. The exciton exhibits significant delocalization along the atomic chain direction (\textit{a}-axis), consistent with the quasi-1D nature of the crystal. In the bulk phase, excitons exhibit relatively smaller binding energies and large radii, spanning multiple unit cells, which is characteristic of Wannier-type excitons. In contrast, the Janus monolayer (see Fig.~\ref{wf}c) exhibits a more localized exciton, extending over 3-4 unit cells. However, in both the exfoliated monolayer and monochain structures, as shown in Fig.~\ref{wf}d and e, the exciton becomes highly confined to a single atomic chain, leading to significantly higher exciton binding energies compared to the bulk. The substantial E$_b$ suggests the formation of strongly bound Frenkel excitons. This progression from weakly bound Wannier-type excitons in bulk BiSeI to strongly confined excitons in its low-dimensional derivatives demonstrates the fundamental impact of reduced dimensionality on excitonic properties.

\subsection*{Conclusion}
We have systematically investigated the electronic and optical properties of bulk, Janus monolayer, exfoliated monolayer and monochain BiSeI using $\textit{ab initio}$ methods based on DFT and many-body approaches. The electronic band structures were computed using the PBE and $\mathit{G_0W_0}$ methods, with the latter providing reliable band gaps. Bulk, Janus monolayer, and exfoliated monochain BiSeI exhibit indirect fundamental band gaps, while the exfoliated monolayer features a direct band gap at the Z point. A significant increase in the fundamental band gap and quasi-particle correction is observed with decreasing dimensionality, attributed to reduced dielectric screening and quantum confinement effects. The fundamental band gaps across all configurations fall within the infrared to the visible range, emphasizing their potential in photovoltaic applications. 

Excitonic properties were analyzed by solving the BSE, revealing that bulk BiSeI hosts spatially delocalized excitons with low binding energies. In contrast, lower dimensional structures exhibit enhanced exciton binding energies due to increased confinement. The Janus monolayer confines excitons to a few unit cells, with a binding energy of 0.34 eV - approximately four times higher than that of bulk BiSeI. In the exfoliated monolayer and monochain structures, excitons are tightly bound within a single atomic chain, leading to significantly higher binding energies. This study provides a comprehensive understanding of the impact of dimensionality on the excitonic properties of BiSeI, highlighting their potential for future optoelectronic and excitonic applications. 

\subsection*{Acknowledgment}
S.M. acknowledge IIT Delhi for the senior research fellowship. S.B. acknowledge financial support from SERB under a core research grant [Grant no. CRG/2019/000647] to set up his High Performance Computing (HPC) facility ``Veena'' at IIT Delhi for computational resources.


\end{document}